\documentclass[11pt,
prd,aps,amssymb,amsmath  ,tightenlines
]{revtex4}

\usepackage{algorithm}
\usepackage{algpseudocode}
\usepackage{graphics}
\usepackage[pdftex]{graphicx}

\usepackage{amsfonts}
\usepackage{amssymb}

\newcommand{\ket}{\rangle}
\newcommand{\bra}{\langle}

\begin{document}

\title{A quantum double-or-nothing game: The  Kelly Criterion for Spins}
\author{ Bernhard K Meister$^{1}$ \& Henry C W Price$^{2}$}
\affiliation{ 
$^{1}$FastEagle Holdings, Vienna, Austria \\ 
$^{2}$Centre for Complexity Science \& Physics Department, Imperial College London, 
London SW7 2AZ, UK \\
}
\date{\today}




\email{bernhard.k.meister@gmail.com, henry.price10@imperial.ac.uk }

\begin{abstract}
\noindent
 A sequence of spin-1/2 particles polarised in one of two possible directions is presented to an experimenter, who can wager in a double-or-nothing game 
  on the outcomes of measurements in freely chosen polarisation directions. Wealth is accrued through astute betting.   As information is gained from the stream of particles, the measurement directions are progressively adjusted, and the portfolio growth rate is raised. 
 The optimal quantum strategy is determined numerically and  shown to differ from the classical strategy, which is associated with the Kelly criterion. 
 The paper contributes to the development  of quantum finance, as aspects of portfolio optimisation are extended  to the quantum realm. 
\end{abstract}



 
\maketitle


\setcounter{section}{0} 
\section{Introduction}
\label{sec:1a}
\noindent
An early application of information theory to gambling and finance can be found  in a 1956 paper by John L. Kelly\cite{Ke1956} $-$ an associate of Claude Shannon. Kelly showed how a logarithmic utility maximising investor should allocate capital between bets with known  winning probabilities  and pay-outs. In the simplest case  the investor bets a fraction of  wealth on the outcome of a biased coin in the `double-or-nothing' game. An outcome of `heads' doubles the gambler's stake, while `tails' loses the stake.  The optimal fraction $-$ called the `Kelly criterion' $-$ is to bet 
$2p-1$ for a coin that comes out heads with probability $p$ $-$ assuming $p\geq 1/2$. 
 This maximises the logarithmic utility,
  which is equivalent to maximising the growth rate of the 
  portfolio. 

Kelly's results are an elaboration of earlier work of Daniel Bernoulli and others, who studied the `St. Petersburg paradox'. The question was how to evaluate a  coin  flipping game, where the pay-out depends on how often 'heads' occurs in succession\footnote{In the play `Rosencrantz and Guildenstern Are Dead' by Tom Stoppard `heads' occurs over eighty times in succession.}. If  'tails' first occurs on the $N$-th throw, then the pay-out is equal to $2^{N}$.   The coin is considered 'fair' and the probability for a having a pay-out $2^N$ is $2^{-N}$.
The paradox was that whilst the expectation value of the bet is not finite the value gamblers were willing to pay for such a pay-out was finite.    One `resolution' of the paradox goes back to Daniel
Bernoulli, who in 1738 suggested that large gains should be
discounted more than smaller gains or losses, since gains and losses are perceived non-linearly due to risk-aversion. In general, returns should be 
adjusted for risk. As the saying goes,  `a bird in the hand is better than two in the bush'.
 Bernoulli suggested as a   value function, which turns gains into   utility,  the logarithm of the gain. This shifts benefits from absolute to relative gains and makes the result independent of the current wealth of the bettor. 
 His derivation uses the equation $dy = k \frac{dx}{x}$,
where $y$ is value or utility, $x$ the wealth and $k$ a constant.
This leads to $y=k \log  x +c$.
The result is that gambles are worth taking,
if the value or utility increases, when participating in the gamble
at the price asked.
 The utility of the bet following Bernoulli's description is then
\begin{eqnarray}
\sum_{i=1}^{\infty} 2^{-i} \log  (2^i)=\log {4}\nonumber
\end{eqnarray}
assuming $k=1$ and  $c=0$.
Therefore, a bettor following Bernoulli's rationale would be willing to pay up to $4$ units to play the
game. A person offering the bet would have to account for possible big losses, which are accounted differently   from big gains, and would not normally arrive at the same price. 

The  work of Kelly was championed by Cover, Ziemba  and others, and found general application in finance. It overlaps with the popular mean-variance portfolio theory introduced by Harry Markowitz. A review of modern developments of the Kelly criterion can be found in MacLean {\it et al.} \cite{Zie2011}, and some applications are in \cite{lv2010,meister2016,meister2022,meister2023}.

Instead of a series of coin flips we consider a sequence of examined spin-1/2 particles. 
Again, there is  a `double-or-nothing' game to wager on, but the pay-outs depend 
  on the results of variable measurements. 
What distinguishes the classical from the quantum case is the added degree of freedom associated with the measurement directions of the quantum particles. Instead of just flipping a coin one can measure the spin particle in an arbitrary direction. The outcome probability is direction dependent. 
This paper highlights some novel aspects of  quantum game theory, quantum gambling and in extension quantum finance.
`Quantum gambling' is likely to have applications in finance as  the quantum scale becomes of practical importance in communication and computation. The importance of hedging against errors that have economic consequences rises. The quantum gambling toy model described in the paper should be of relevance for analysing such situations. 
The conclusion provides further information.


A summary of the rest of the paper  follows.  The next section introduces gambling with spin particles, followed by a section on the   quantum version of the coin flipping game. 
The  subsequent section give a heuristic description of the optimal strategies. Section five   delves into numerical calculations.  In a sub-section the special case of prior $1/2$ is considered. The penultimate section covers the optimal strategy expressed  as 'contours of equivalence', and a conclusion rounds off the paper.

\section{Gambling with Spin Particles}
\noindent
This section describes how one can `quantum gamble'  with spin particles. 
A gambler, or to use the more courteous term: investor\footnote{W. F. `Blackie'  Sherrod: ``If you bet on a horse, that's gambling. If you bet you can make three spades, that's entertainment. If you bet cotton will go up three points, that's business. See the difference?"
}, is presented with a sequence of quantum spin-1/2 particles. Unbeknown to the investor is the polarisation of the particles, which are either all prepared 
 in state $\rho$ or state $\sigma$. The investor is only informed that the probability for $\rho$ is $\xi$ and for $\sigma$ is $1-\xi$. 
The investor is further told that a `double-or-nothing' game is linked to  measurements of the particles. The investor can bet any fraction\footnote{Leverage is not considered.}  of owned assets  on the outcome of the measurement.  If the measurement outcome in the direction of  choice is spin-up, then the investor's stake is doubled. If, on the other hand, the outcome is spin-down, the  stake is forfeited. 

If only one bet is considered, then the straightforward aim is to maximize  the winning probability, which fixes the measurement direction. If a sequence of bets is to be considered, another factor influences the choices, since judicious measurements are accompanied by a gain of information, and allow progressively a more and more accurate determination of the polarisation direction of the particles. 
Balancing short-term winning probability with information gain and increased profitability is novel and reflects the quantum nature of the problem. In the next section, some relevant calculations for spin gambling are presented.

\section{The `double-or-nothing' game with spin-1/2 particles.}
\label{sec:2a}
\noindent
This section gives further details   of the spin-1/2 gambling game.
Notation conducive to the problem at hand is next introduced. 
The candidate states are  $|\psi\ket=\sin{\gamma}|0\ket +\cos{\gamma}|1\ket $ and  $|\phi\ket=\sin{\beta} |0\ket + \cos{\beta}|1\ket $, more conveniently written as $|\phi\ket = \cos{\delta} |\psi\ket  + \sin{\delta} |\psi_{\bot}\ket  $  with \begin{eqnarray}
 \cos^2{\delta}=( \sin{\gamma} \sin{\beta} + \cos{\gamma} \cos{\beta})^2= \cos^2(\gamma-\beta).\nonumber
 \end{eqnarray} 
 To simplify, we set $\gamma:=0$, and the remaining relevant angle is $\delta$. 
 The  density matrices are $\rho= |\phi\ket \bra \phi | $  and $\sigma= |\psi\ket \bra \psi | $  with the respective priors  of $\xi$ and $1-\xi$. The derivation of the optimal informational strategy can be found in Brody {\it et al.}\cite{br1996} for the case of two polarisation directions. If one considers additional polarisation directions, the optimal strategy has only been obtained numerically.
 
  The tunable inputs are then reduced to $\delta$, $\xi$ and the number of available particles $N$. The index $i$ runs from $1$ to $N$. The initial wealth $W_1$ is $1$ without loss of generality, since the wealth axis\footnote{Later in the paper we do not fix the initial  but the final wealth. If for a prior $\xi_0$ one wants a particular initial wealth $W_0$ one can re-scale the wealth axis and continue to use the already calculated contour lines.} can be arbitrarily scaled. 
 The  optimal measurement angle to maximise the information gain is 
  \begin{eqnarray}
  \phi 
  =\tan^{-1}\Big(\frac{(\xi-1) \sin(\delta)}{\xi - (1-\xi)\cos(\delta)}\Big),
 \nonumber
\end{eqnarray} 
and the optimal portfolio growth angle ${\hat\phi}$ is $\phi$ shifted by $\pi/4$,
i.e. ${\hat\phi} 
=\phi 
+\pi/4$, if and only if $\xi=1/2$.
For each of the $N$ particles gambled on and analysed a separate $\alpha_i$ is chosen.
The probability 
of a spin-up outcome for such a measurement in the $\alpha_i$ direction is 
\begin{eqnarray}
p^{up}_{i+1}=\xi_i \cos^2(\alpha_i)+ (1-\xi_i)   \cos^2 (\delta - \alpha_i)
.\nonumber
\end{eqnarray} 
The updating of the prior leads for spin-up outcomes to
\begin{eqnarray}
\xi^{up}_{i+1}=\frac{\xi_i \cos^2(\alpha_i)}{\xi_i \cos^2(\alpha_i)+ (1-\xi_i)   \cos^2(\delta - \alpha_i)}
,\nonumber
\end{eqnarray} 
and for spin-down outcomes to
\begin{eqnarray}
\xi^{down}_{i+1}=\frac{\xi_i \big(1-\cos^2(\alpha_i)\big)}{\xi_i \big(1-\cos^2(\alpha_i)\big)+ (1-\xi_i) \big(1-  \cos^2(\delta - \alpha_i)\big)}
.\nonumber
\end{eqnarray} 
The optimal investment fraction for a chosen $\alpha_i$ is $f_i=2 p^{up}_i-1$, if $p^{up}_i\geq 1/2$ and otherwise $\alpha_i $ is shifted by $\pi/2$ replacing  $p^{up}_i$ by $1-p^{up}_i$.
The updating of the wealth process leads, in the case of a spin-up outcome, to
\begin{eqnarray}
W_{i+1}=W_{i}(1 + f_i) ,\nonumber
\end{eqnarray} 
and, in the case of a spin-down outcome, to
\begin{eqnarray}
W_{i+1}=W_{i}(1 - f_i).\nonumber
\end{eqnarray} 
If one now simulates a chain of measurements in the $\alpha_i$ 
directions, then one gets a series of 
spin-up and spin-down outcomes with probabilities  $p^{up}_i$ and $1-p^{up}_i$.
Let's represent a spin-up outcome by $+$ and a spin-down outcome by $-$.
The outcome sequence would then look like $+-+++-...+-++++++$ with progressively more $+$'s as one edges the prior away from $1/2$. 
The optimisation algorithm will adjust $\alpha_i$ to maximise the probability-weighted logarithmic utility of $W_N$
 \begin{eqnarray}
 \max_{all \,\,\,possible\,\,\, \alpha_i \,\,\,sequences}\sum_{all \,\,\,the\,\,\ 2^N \,\,\,+/- \,\,\,sequences} p(...)\log W_N(...).
\nonumber
\end{eqnarray} 
The different outcome sequences have probabilities $p(...)$, i.e. all the way from $p(+ \,... \,+)$ to $p(- \,...\, -)$,  associated with them. Their probability is a product of $N$ terms of the form $p^{up}_{i}$'s and $p^{down}_{i}:=1-p^{up}_{i}$, i.e. $i$ running from $1$ to $N$.
Next, a section on the optimal strategy.

 
 \section{The Optimal Strategy: A Heuristic Description}
 \label{sec:3a}
\noindent
 In this section, a heuristic description of the the optimal strategy is given. It requires the backward solution of  an optimization problem, similar to the evaluation of a European option on a binomial tree. 
 
 If only one particle remains to be measured, then maximization of the spin-up probability is key, since any `information gain' cannot be exploited in the future. 
 Therefore, one maximises in the last round solely portfolio growth. This determines the angle $\alpha_N$  and with it the spin-up probability, i.e. 
 \begin{eqnarray}
 0 &= &\partial_{\alpha_N} \Big( \xi\cos^2(\alpha_N) +(1-\xi)\cos^2(\alpha_N-\delta) \Big)    \nonumber \\
 0&=& \xi \cos(\alpha_N) \sin(\alpha_N) +(1-\xi) \cos(\alpha_N-\delta) \sin(\alpha_N-\delta) \nonumber \\
  0&=&\tan^2(\alpha_N) \sin(\delta) \cos(\delta)+\tan (\alpha_N)\Big(\frac{\xi}{1-\xi}+1-2\sin^2(\delta)\Big) - \sin(\delta) \cos(\delta)  \nonumber \\
&& \Longrightarrow  \alpha_N =\tan^{-1}\Big(- \frac{A}{2}\pm\frac{1}{2}\sqrt{A^2+4}\,\, \Big) 
\nonumber
 \end{eqnarray}
 with $A=\frac{\frac{\xi}{1-\xi}+1 - 2 \sin^2(\delta)}{\sin(\delta)\cos(\delta)}$. 
  The `Kelly criterion' allows then determination of the optimal fraction, and the resulting wealth  (and $\xi$) can  be calculated for both measurement outcomes, and with it  the achievable probability-weighted utility. 
 
 Next we move one step back. The angle $\alpha_{N-1}$ at step $N-1$ updates in a distinct way for the two possible measurement outcomes both the wealth and the prior. By maximising over the range of allowed measurement angles $\alpha_{N-1}$ one can determine the maximal achievable utility as  the probability-weighted  sum of the utility at the final step. This is then the utility associated with that particular point  in the two dimensional wealth \& $\xi$ space. 
 A similar mechanism works for earlier steps all the way back to step one. 
 
 
The optimal strategy is then  given by the contour map of equivalent utility lines at each step over the wealth and $\xi$ space. Figures 1-6 and 9-10 show  examples of contour lines for the  polarisation angle $\delta$ of $7.5^\circ, \,\,30^\circ, \,60^\circ$ and $90^\circ$. Due to limitations of the mesh some boundary effects can be seen  in figures 5 $\&$ 6.    Each angle comes with a pair of plots, i.e. with and  without the wealth presented on a logarithmic scale.
The points connected by the $k$-th contour 
line 
yield the same final wealth utility at step $N$. Any logarithmic utility maximising bettor should be
indifferent between any of the points on these lines.
The endpoints of the $k$-th contour line correspond  to $\xi=1$  or $\xi=0$ and wealth of $W_N/(2^(N-k))$, i.e. a doubling of wealth until one reaches $W_N$. The other points on the same contour line have higher wealth but a $\xi$ closer to $1/2$, and therefore lower optimal winning probability. Besides the contour graphs the heat maps (see figures 7-8) are also of interest, which show that utility is strongly linked to the remaining number of steps as well as wealth $W$, but less to the prior $\xi$. 
Next, a description of the optimal strategy, derived by numerical means. 

\section{Numerical Simulation: Algorithm and Pseudo Code}
\noindent
The numerical simulation is the topic of this section. The algorithm works backwards, and defines the utility surface at each step as a function dependent on the wealth value $W$ \& prior $\xi$.
At the final step $N$, all points on the straight line contour have a  fixed wealth  $W_N$ but arbitrary value for $\xi$. This can be represented as a vector $\Big(W_N,\xi_N, N, \log_2(W_N)\Big)$. The utility surface for the previous step is then defined, and general equations for calculating $\Big(U_k(W,\xi)\Big)$ are taken from section III. The utility value for each step is calculated through an iterative process. 
\noindent
The pseudo-code below outlines the computational method for calculating the utility surface.
The result is a multidimensional array of utility values, with non-grid-point utilities  evaluated by interpolation. 
\begin{algorithm}[h!]
\caption{Quantum Kelly Optimization across N steps}
\begin{algorithmic}[1]
\State Define $\alpha_{\text{range}}$ from 0 to $\frac{\pi}{2}$
\State Define $W_{\text{range}}$ with logarithmically spaced values between 0 and 2
\State Define $\xi_{\text{range}}$ with logarithmically spaced values between 0 and 0.5
\State Define $\delta$ (adjustable parameter)
\State Initialize utility array $U$ with zeros
\State Compute $U[N-1, :, :] = \log_{2}(W_{\text{range}})$ \Comment{Initial value using log2}

\For{$k = N-2$ to $0$}
    \State Create interpolation for current stage based on $W_{\text{range}}$ and $\xi_{\text{range}}$
    \For{each $W$ in $W_{\text{range}}$}
        \For{each $\xi$ in $\xi_{\text{range}}$}
            \State Compute $p_{+}$, $p_{-}$ based on $\xi$, $\alpha$, $\delta$
            \State Compute $\xi_{+}$, $\xi_{-}$ based on $\xi$, $\alpha$, $\delta$
            \State Compute $W_{+}$, $W_{-}$ based on $W$, $p_{+}$
            \State Interpolate values $U_{+}$, $U_{-}$ for given $W_{+}$, $\xi_{+}$, $W_{-}$, $\xi_{-}$
            \State Define utility function using the above values
            \State Optimize $\alpha$ to find $\alpha_{\star}$ that maximizes utility
            \State Update $U[k, i, j]$ with optimized utility value \Comment{Iterative update of U using probability weighted sum of previous step utilities}
        \EndFor
    \EndFor
\EndFor
\end{algorithmic}
\end{algorithm}
\noindent
Visualization is achieved by extracting contours from the array. 

\noindent

\noindent
The simulation was written in Python;  \texttt{NumPy} was used  for numerical operations; \texttt{SciPy}  \texttt{optimized} and \texttt{interpolated} the modules; and \texttt{Matplotlib} was employed for visualization.

\section{Contours of Equivalence }
\noindent
In this section a technique for visualising the optimal gambling strategy is introduced. It relies on 
 finding points in the two dimensional wealth and prior space that  lead to the same final `utility' outcome. 
These sets of equal `utility' points form contour lines. Each line differs by the number of remaining particles to be measured and, consequently, bets to be made. 

\begin{figure}
  \centering
  \begin{minipage}{0.47\textwidth}
    \centering
    \includegraphics[width=\textwidth/2]
    {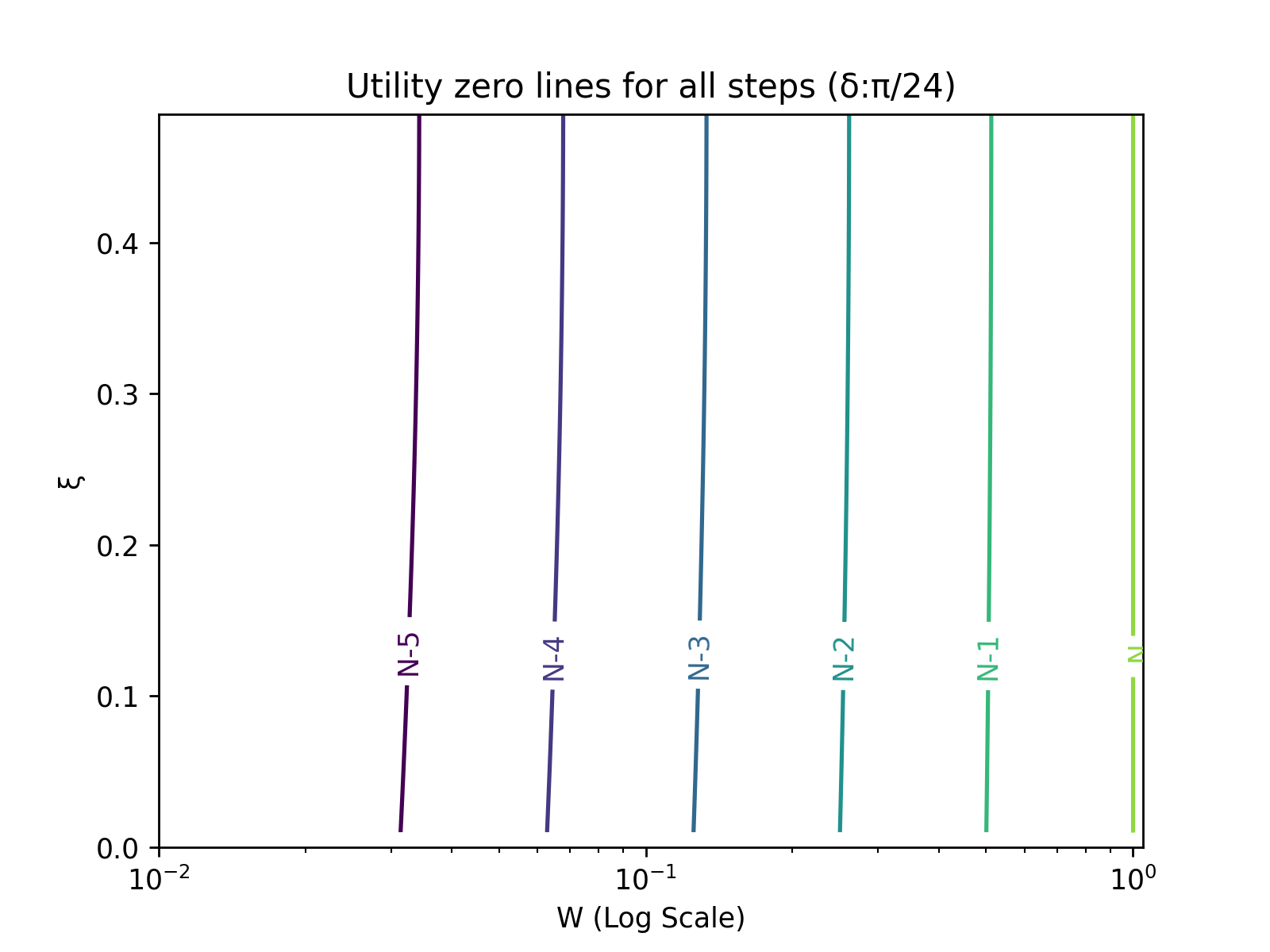}
    \caption{Plot of contour lines for same Utility at different steps for $\delta=7.5^{\circ}$ using a logarithmic scale on the x-axis for $W$}
    \label{fig:label1}
  \end{minipage}\hfill
  \begin{minipage}{0.47\textwidth}
    \centering
    \includegraphics[width=\textwidth/2]
    {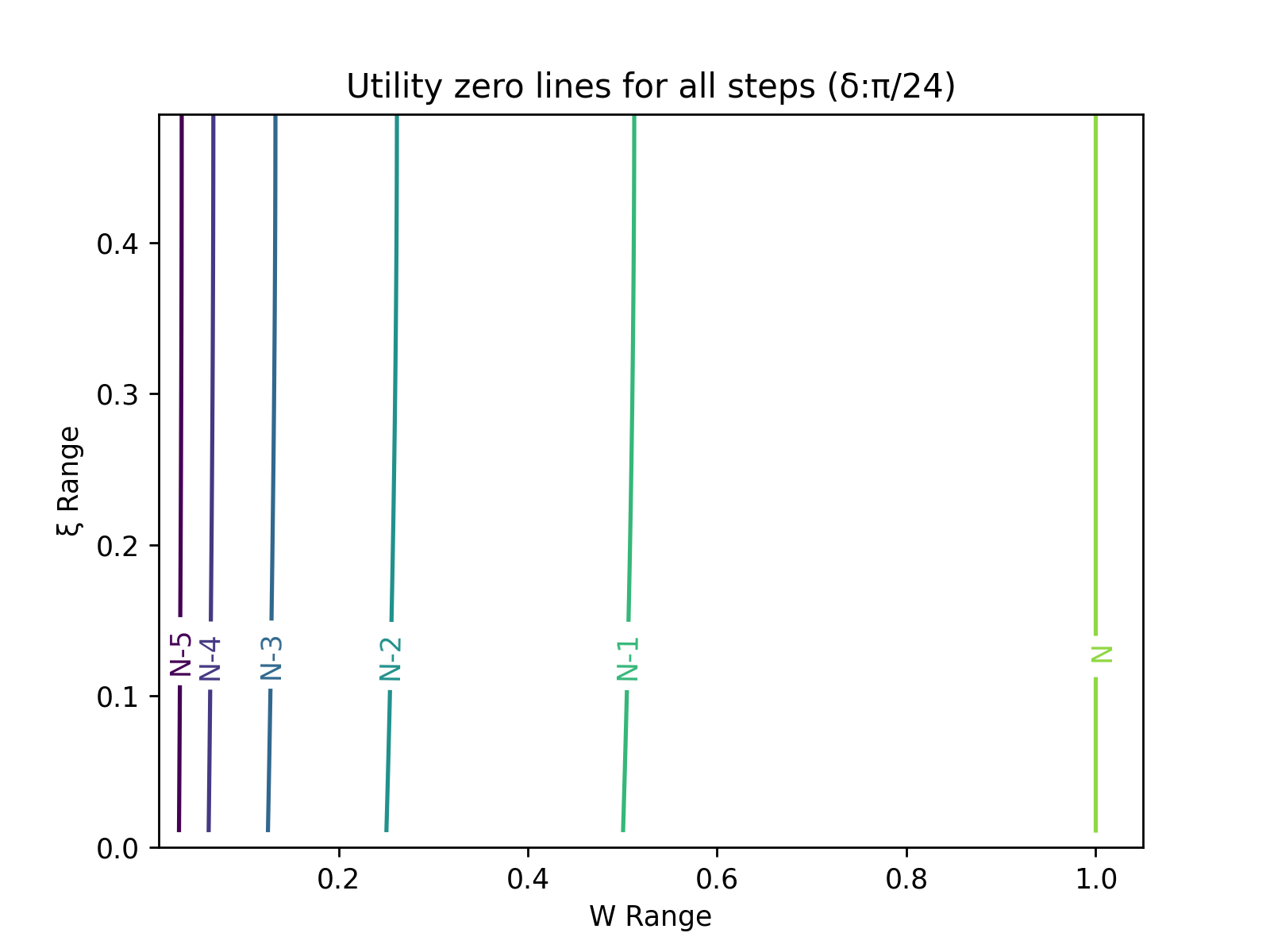}
    \caption{Plot of contour lines for same Utility at different steps for $\delta=7.5^{\circ}$}
    \label{fig:label2}
  \end{minipage}
\end{figure}

\begin{figure}[h]
  \centering
  \begin{minipage}{0.47\textwidth}
    \centering
    \includegraphics[width=\textwidth/2]
    {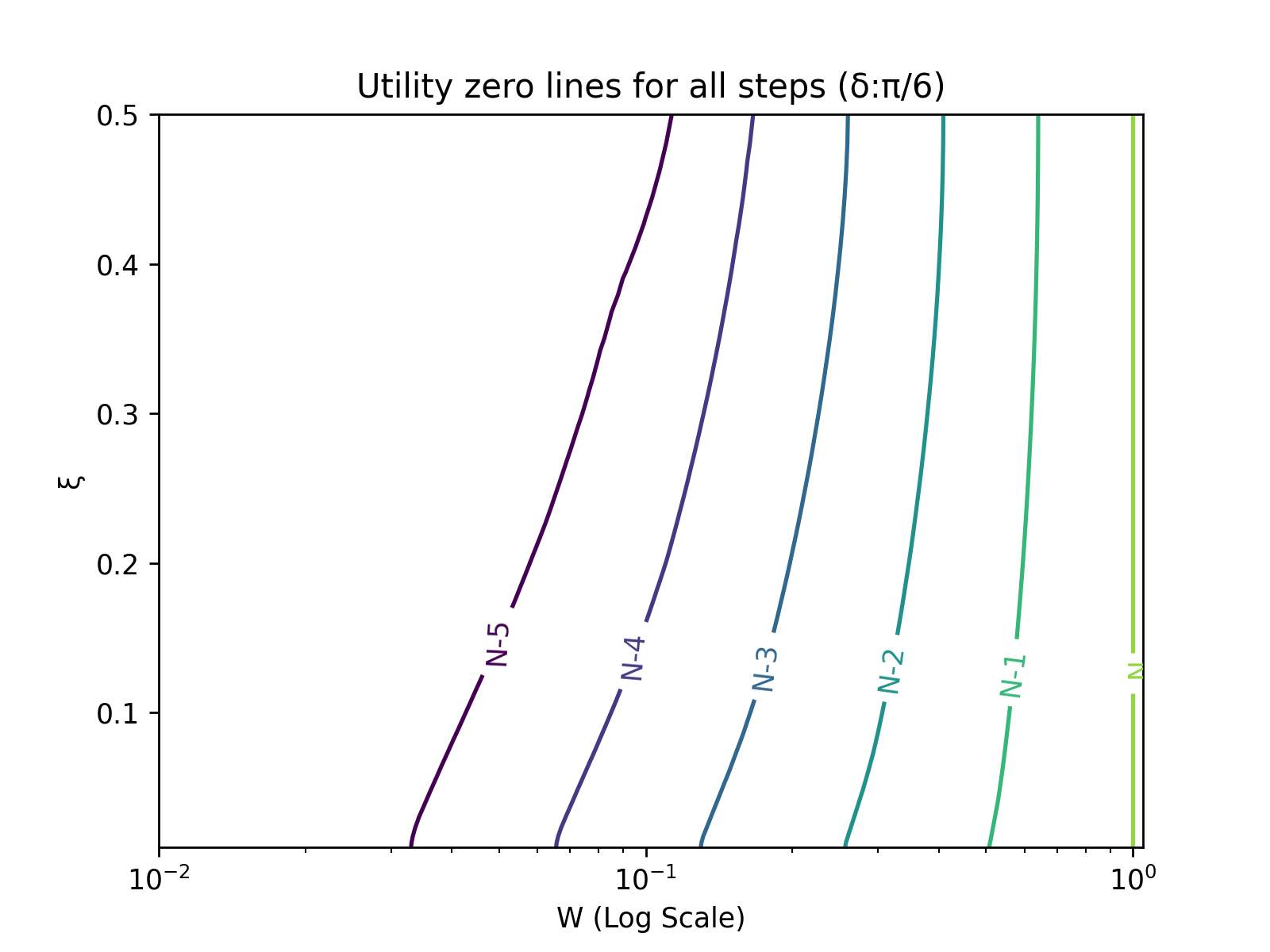}
    \caption{Plot of contour lines for same Utility at different steps for $\delta=30^{\circ}$ using a logarithmic scale on the x-axis for $W$}
    \label{fig:label1}
  \end{minipage}\hfill
  \begin{minipage}{0.47\textwidth}
    \centering
    \includegraphics[width=\textwidth/2]
    {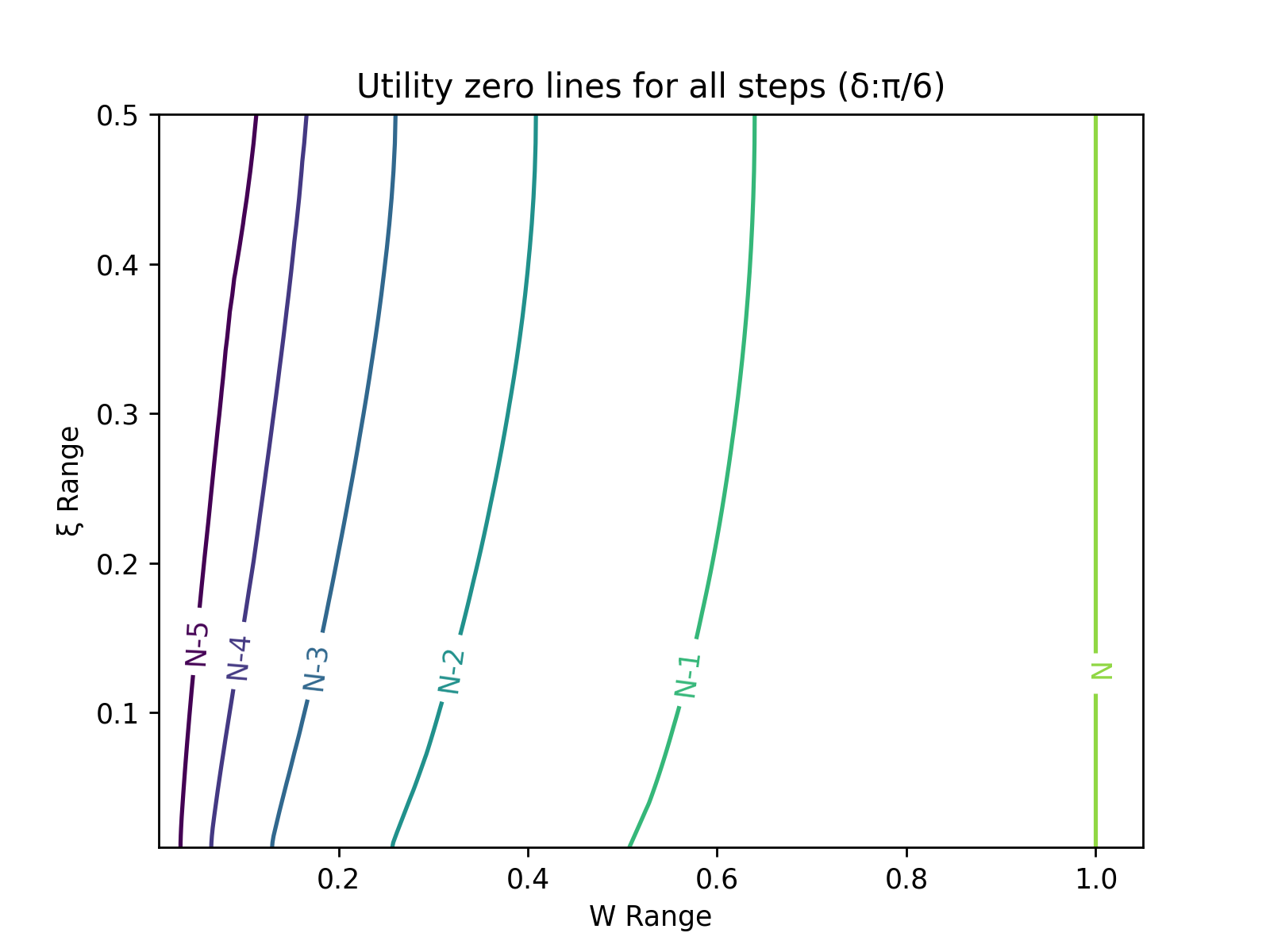}
    \caption{Plot of contour lines for same Utility at different steps for $\delta=30^{\circ}$}
    \label{fig:label2}
  \end{minipage}
\end{figure}

\begin{figure}[h]
  \centering
  \begin{minipage}{0.47\textwidth}
    \centering
    \includegraphics[width=\textwidth/2]
    {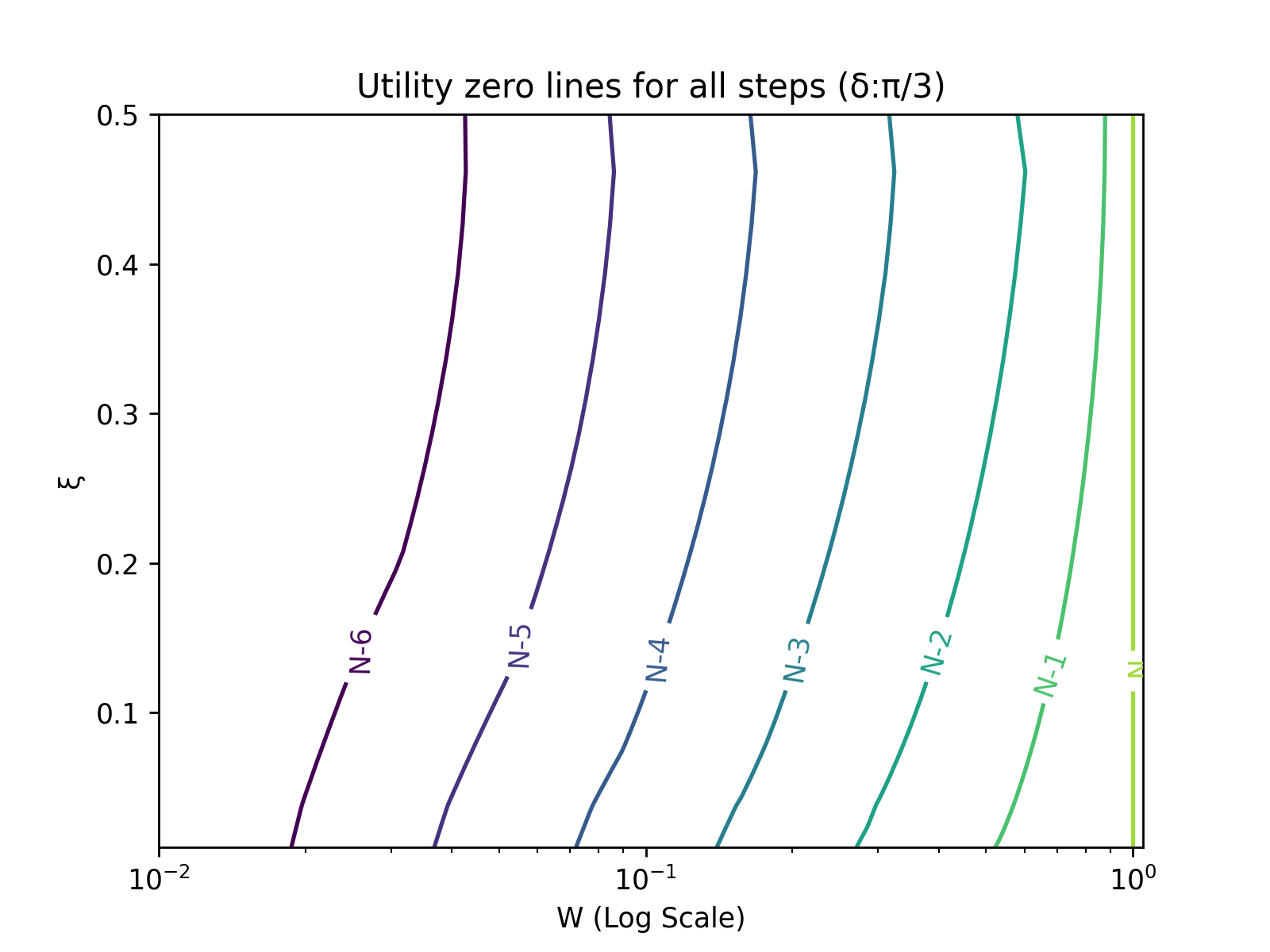}
    \caption{Plot of contour lines for same Utility at different steps for $\delta=60^{\circ}$ using a logarithmic scale on the x-axis for $W$}
    \label{fig:label1}
  \end{minipage}\hfill
  \begin{minipage}{0.47\textwidth}
    \centering
    \includegraphics[width=\textwidth/2]
    {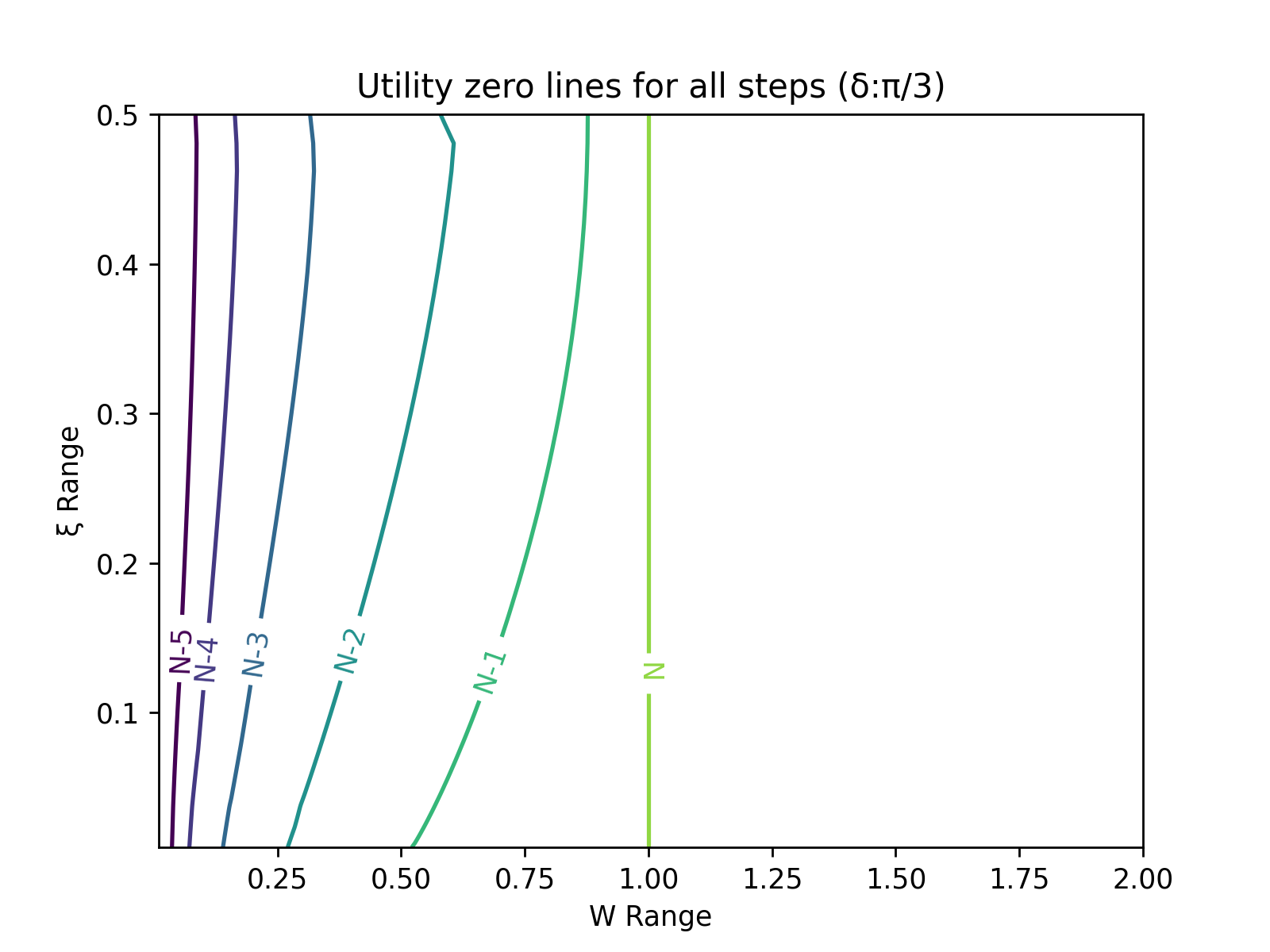}
    \caption{Plot of contour lines for same Utility at different steps for $\delta=60^{\circ}$}
    \label{fig:label2}
  \end{minipage}
\end{figure}

These contour lines all have easily calculated levels of wealth at the $k$-th step from the end for  extreme values of the prior. These anchor points for the prior  $\xi$ at $0$ and $1$ are associated with a wealth of $2^{k-N}W_N$. There is a  simple strategy that moves from these points to the  final contour. Since the probability of winning is one, the gambler bets everything and doubles the money at every stage. 
Between these two extremes lies a range of points, which require numerical evaluation. 

\begin{figure}[h]
  \centering
  \begin{minipage}{0.47\textwidth}
    \centering
    \includegraphics[width=\textwidth/2]
    {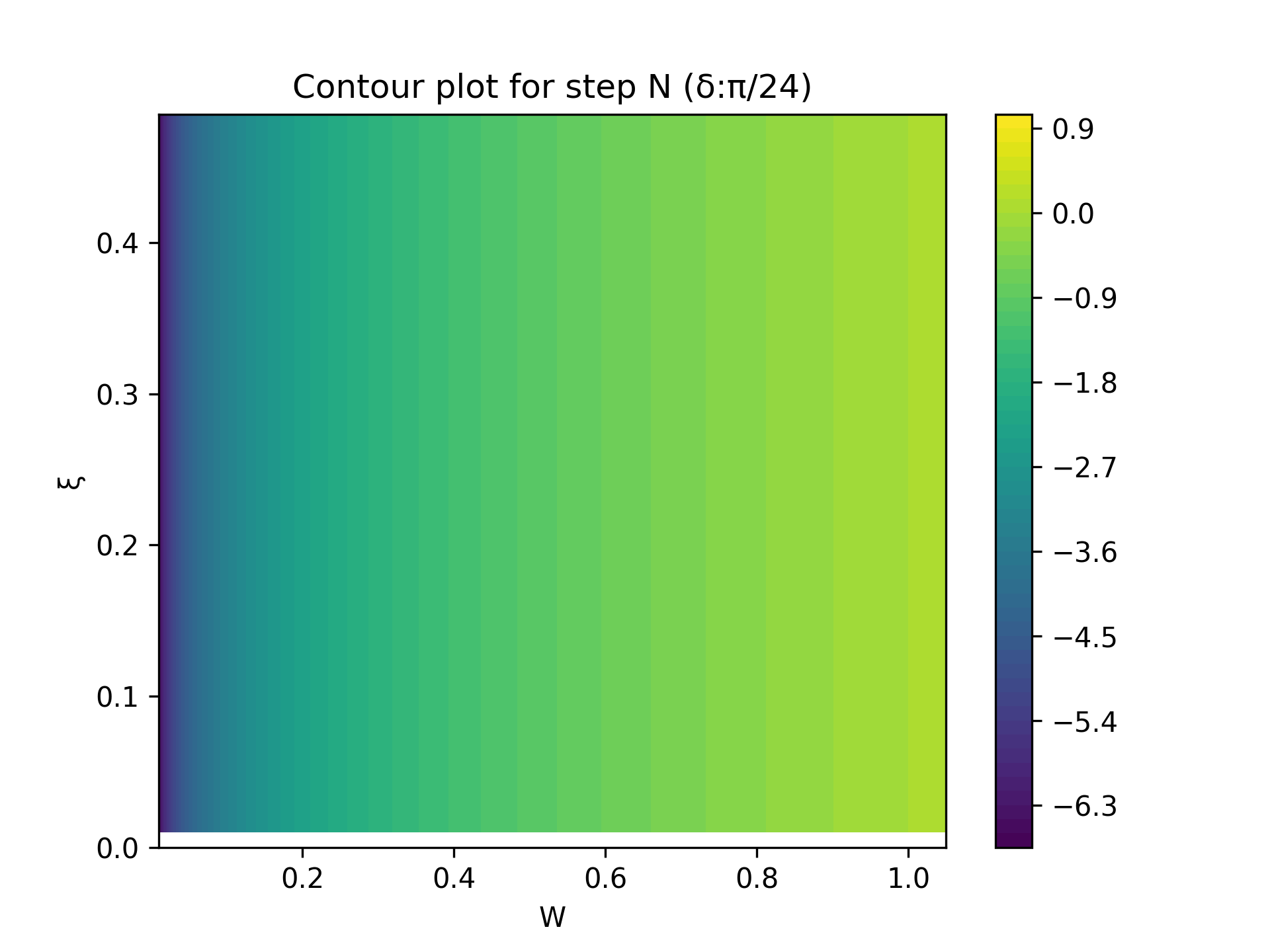}
    \caption{Heat map of Utility at the final step}
    \label{fig:label1}
  \end{minipage}\hfill
  \begin{minipage}{0.47\textwidth}
    \centering
    \includegraphics[width=\textwidth/2]
    {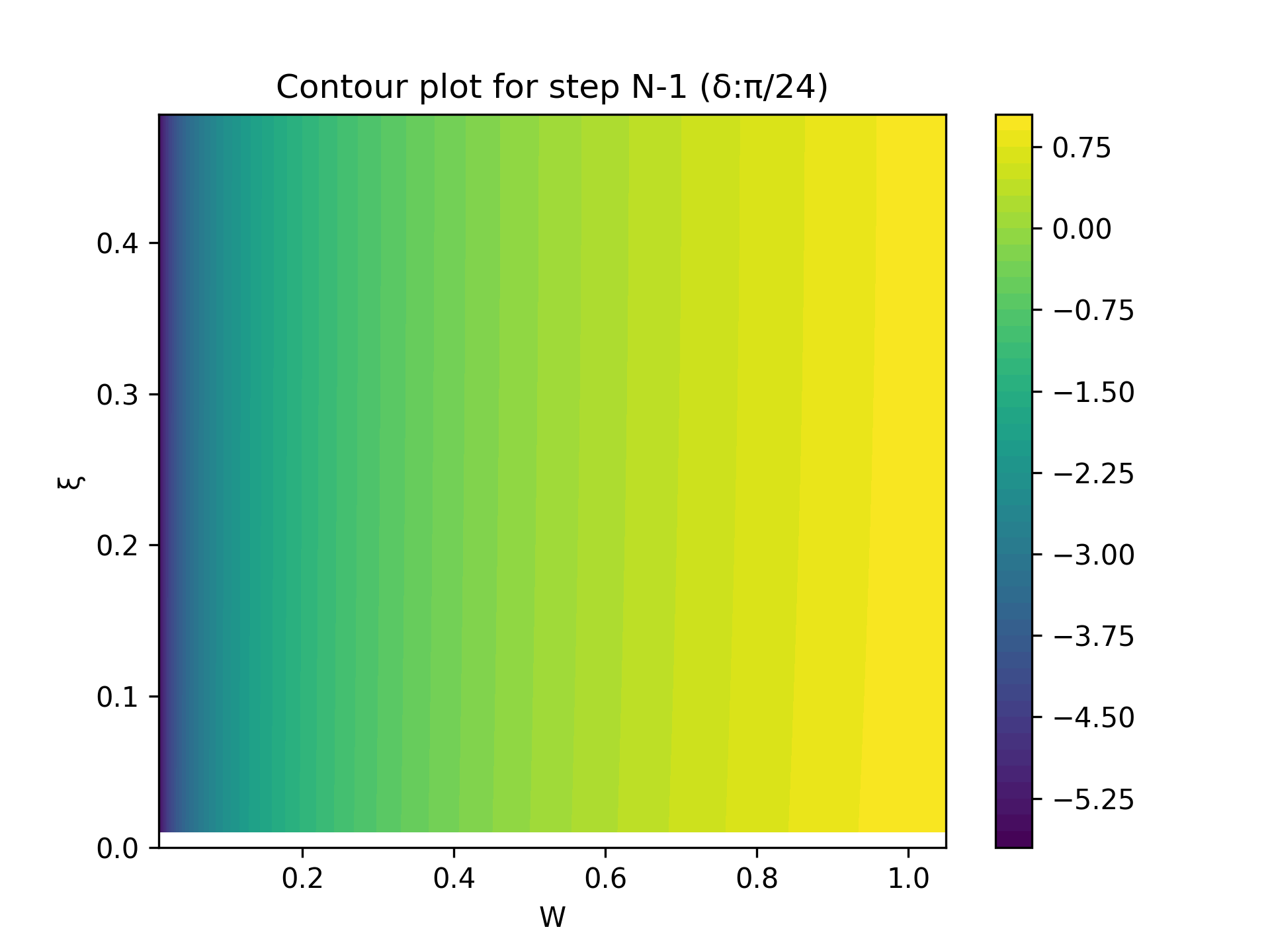}
    \caption{Heat map of Utility at the second to final step}
    \label{fig:label2}
  \end{minipage}
\end{figure}


\subsection{The curious case of prior one half: growth without information gain or information gain without growth}

\begin{figure}[h]
  \centering
  \begin{minipage}{0.47\textwidth}
    \centering
    \includegraphics[width=\textwidth/2]
    {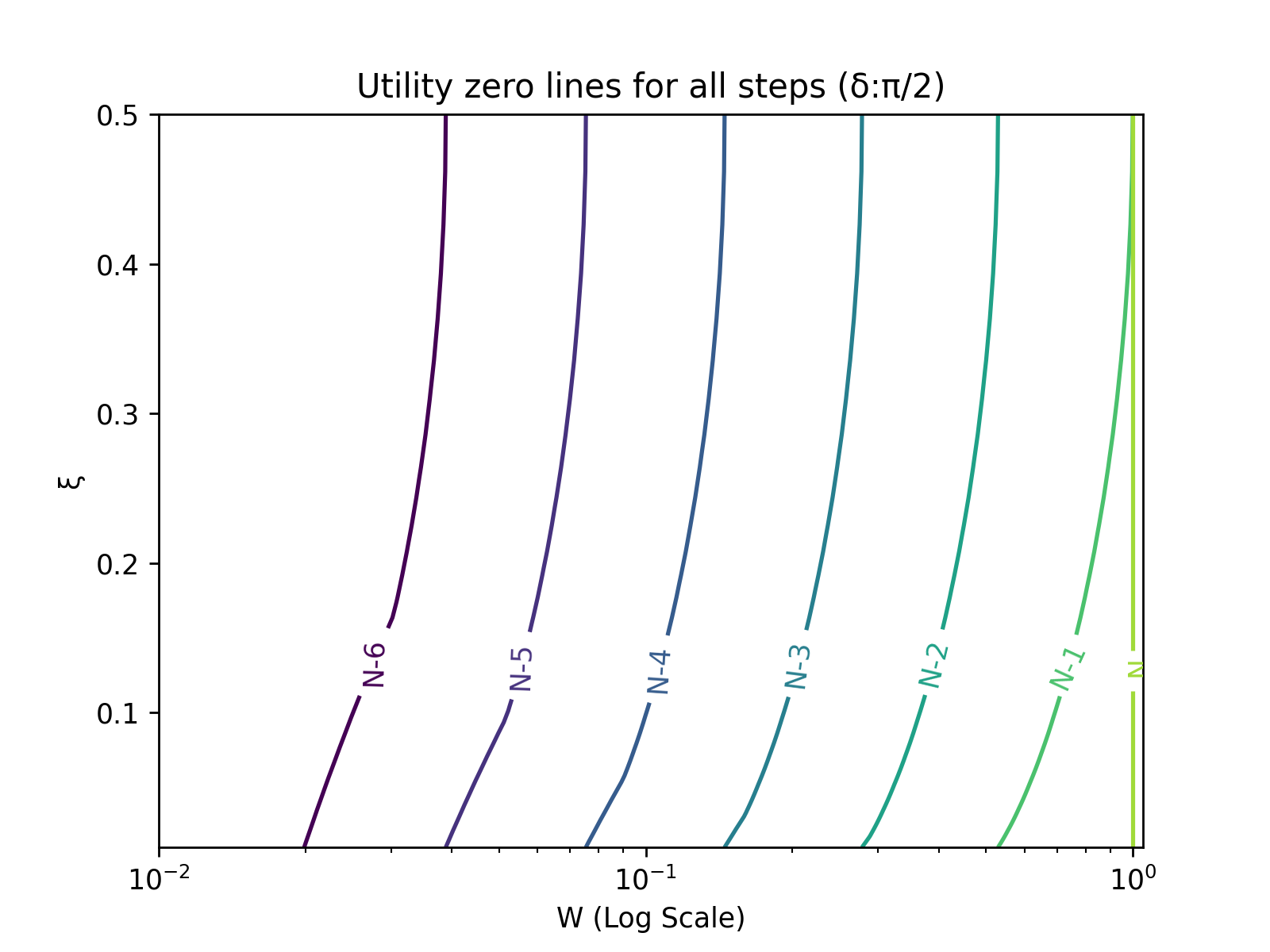}
    \caption{Plot of contour lines for same Utility at different steps for $\delta=90^{\circ}$ using a logarithmic scale on the x-axis}
    \label{fig:label1}
  \end{minipage}\hfill
  \begin{minipage}{0.47\textwidth}
    \centering
    \includegraphics[width=\textwidth/2]
    {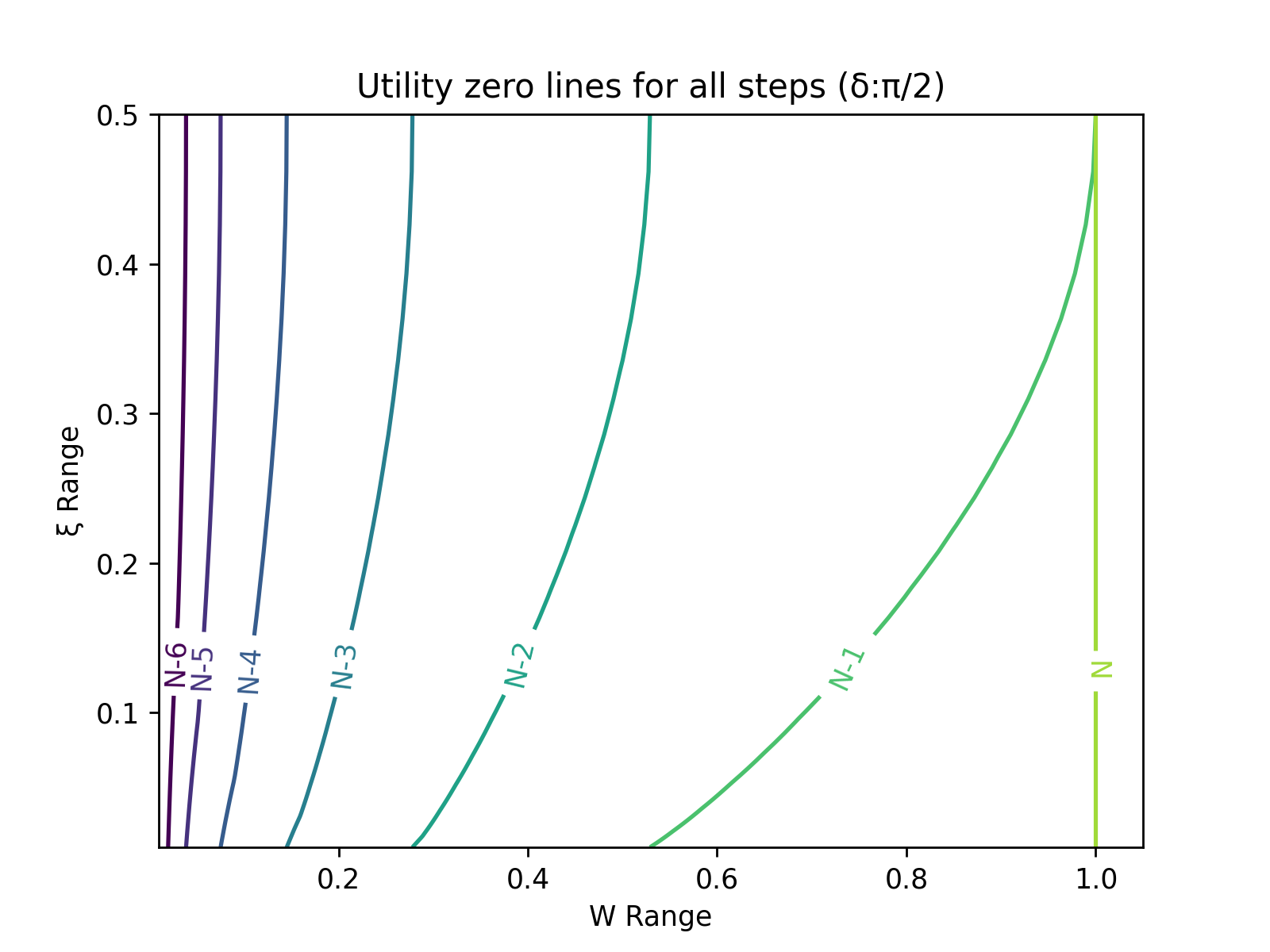}
    \caption{Plot of contour lines for same Utility at different steps for $\delta=90^{\circ}$ using a standard scale on the x-axis 
    }
    \label{fig:label2}
  \end{minipage}
\end{figure}

\label{sec:3a}
\noindent
In this subsection, the special case of prior one-half is considered. This choice
simplifies formulae.
The optimal angle for information gain was given in section III  and differs  in this special case from the one-round maximal growth angle given in section IV by $\pi/4$.
The maximal information gain direction entails a betting fraction of zero, whereas the maximal growth measurement direction leaves $\xi$ unchanged.  The two extremal strategies can be said to be  maximally incompatible for $\xi=1/2$. 

\section{Conclusion}
\label{sec:3a}
\noindent
The paper investigated a quantum betting game and showed how an optimal strategy can be established. Some differences between the classical and quantum case were discussed, and some strategies were compared.
Simple extremal strategies were most easy to compare. One could for example maximise `information gain' or `portfolio growth'. As we have shown, the optimal strategy lies somewhere in between and depends on $\delta$,  $\xi$ and the number of steps $N$. If $\xi$ is either zero or one, then information gain is impossible and maximising short and long-term growth becomes synonymous. 
A special case is $\xi$ equals one-half, where an optimal `information gain'  measurement entails zero `portfolio growth', and an optimal `portfolio growth' measurement  entails zero `information gain'.

Applications of `quantum gambling' are maybe currently few and far between in finance or the wider  world outside laboratories. However, besides the ability to probe and test  quantum mechanics, one could envisage scenarios where the outcomes of quantum-scale events play a role in  decision-making. 
In this case, quantum game theory and applications like the quantum gambling problem discussed in the paper are of  interest.

One can further find applications in quantum computing and information theory. 
Imagine a quantum communication channel, where a misinterpretation of the message leads to a financial loss. The evaluation and insurance against such faults can be formulated in certain cases as a quantum gambling problem of the type discussed in the paper. 


\noindent
One of the authors  - B.K.M. -    thanks D.C. Brody and L.P. Hughston for  stimulating discussions.

\end{document}